\documentstyle[aps,prl,twocolumn,epsf]{revtex}

\begin{document}
\title{Quantum L\'evy Processes and Fractional Kinetics}
\author{Dimitri Kusnezov$^1$, Aurel Bulgac$^2$ and Giu Do Dang$^3$}
\address{$^1$ Center for Theoretical Physics,
Sloane Physics Laboratory,
Yale University, New Haven, CT 06520-8120 \\
$^2$ Department of Physics, University of Washington,
Seattle, WA 98195--1560\\
$^3$ Laboratoire de Physique Th\'{e}orique et Hautes Energies,
Universit\'{e} de Paris--Sud, B\^{a}t. 211, 91405 Orsay, FRANCE}
\date{\today }
\maketitle

\begin{abstract}
Exotic stochastic processes are shown to emerge in the quantum evolution of
complex systems. Using influence function techniques, we consider the  dynamics
of a system coupled to a chaotic subsystem described through  random matrix
theory. We find that the reduced density matrix  can display dynamics given by
L\'evy stable laws. The classical limit of these dynamics can be related to
fractional kinetic equations. In particular we derive a fractional  extension
of Kramers equation.
\end{abstract}

\draft


\vspace{0.5cm}

\narrowtext

Whether one studies deterministic Hamiltonian or dissipative systems, one finds
that transport in chaotic systems often resembles some type of stochastic
process.  The dynamics of such systems leads to a rich spectrum of behaviors,
from enhanced diffusion such as tracer diffusion in flows and turbulent
diffusion in the atmosphere, to dispersive diffusion\cite{szk}. Much effort
has been spent in recent years to understand such classical stochastic
processes in chaotic systems, leading to the development of approaches ranging
from fractional kinetic equations\cite{fog,zas,fp}, L\'evy flights\cite{mon} to
random walks in random environments\cite{mon,bouch} and  stochastic
webs\cite{zas1}. One of the common features to all of these is the use of
L\'evy stable laws\cite{kzs}.  It was shown  by L\'evy\cite{levy}, in studies
of extensions of the central limit theorem, that a continuous class of
non--gaussian processes satisfy the same fundamental equation that gives rise
to the theory of gaussian processes, namely the
Chapman--Kolmogorov equation for the conditional probability
$P(q,q',t)$:
\begin{equation}
 P(q,q';t) = \int dq'' P(q,q'',t-t'') P(q'',q',t'').
\end{equation}
The standard solution,
$P(q,q',t)= \exp(-(q-q')^2/4Dt)/(4\pi Dt)^{3/2}  $,
gives rise  to the gaussian processes
and the usual form of the Fokker--Planck equation. The general solutions of
L\'evy provide a way to generalize Brownian motion.

The non--gaussian processes which satisfy (1)
are known as L\'evy stable laws, and have the form:
\begin{equation}
P(q,t)={\cal L}_\alpha ^A(q)=\frac 1{2\pi }\int \exp\left\{ikq-A|k|^\alpha
\right\}dk
\end{equation}
where $0<\alpha \leq 2$ and $A\propto t$.
The case $\alpha =2$ corresponds to gaussian
processes. The L\'evy distributions ${\cal L}_\alpha ^A(q)$
satisfy the scaling relation:
\begin{equation}
{\cal L}_\alpha ^A(q)=A^{-1/\alpha }{\cal L}_\alpha ^1(qA^{-1/\alpha }),
\end{equation}
where for $A=1$ we drop the superscript: ${\cal L}_\alpha ^1(x)={\cal L}_\alpha
(x)$.  For $\alpha <2$, these distributions are characterized by infinite
second moments, as one can easily infer from the asymptotic behavior
for $q \rightarrow \pm \infty $ \cite{mon},
\begin{equation}
{\cal L}_\alpha ^A(q) \approx \Gamma (\alpha ) \sin \frac{\pi \alpha }{2}
\frac{\alpha A}{|q|^{\alpha +1 }}.
\end{equation}
These non--gaussian processes can be related to anomalous transport in a
variety of (classical) physical systems\cite{bouch}, as well as to 
classically chaotic systems.  We have recently
shown that turbulent diffusion can also arise in the time evolution of
complex quantum systems\cite{kbd}. Here we
find that a general form of quantum chaotic backgrounds can
give rise to quantum evolution characterized by L\'evy distributions. Further,
we can connect, in the semi--classical limit, such processes to fractional
kinetic theory, which was initially introduced as a phenomenological
approach to classical anomalous diffusion.

We would like to study the problem of a particle coupled to a chaotic
environment, quantum mechanically. It has been realized in recent years that
the quantum counterpart of chaos is random matrix theory. For systems with
time--reversal symmetry, the random matrices are real--symmetric.
In this Letter we will examine the class of quantum dynamic processes, which
can be realized through the interaction of a particle with a random matrix
background. In contrast to the Caldeira--Leggett approach\cite{cald},
we  assume
from the outset that the background is chaotic, and not necessarily
thermal.
We denote the coordinates of the background by $(x,p)$ and that of the test
particle by $(X,P)$. The Hamiltonian for the background plus interaction
is taken to have the following  form:
\begin{equation}
H_b=h_0(x,p)+h_1(X,x,p).
\end{equation}
In the basis of (many--body) eigenstates of $h_0$, $h_0\mid n\rangle
=\varepsilon _n\mid  n\rangle $ ($n=1,...,N$), we define the matrix of $H_b$ as
\begin{equation}
\lbrack H_b]_{ij}=\varepsilon _i\delta _{ij}+[h_1(X)]_{ij}.
\end{equation}
It is convenient for calculations to choose
an average level density as $\rho(\varepsilon)=\rho_0\exp(\beta
\varepsilon)$. For a background with constant average level density, $\beta=0$,
while for a general many body system, $\beta > 0$.
The chaotic properties of the interaction of the background with the
particle are incorporated into the correlation function (second cumulant):
\begin{equation}\label{eq:secmom}
\langle\!\langle [h_1(X)]_{ij}[h_1(Y)]_{kl}\rangle\!\rangle
={\cal G}_{ij}(X-Y)\Delta_{ijkl}.
\end{equation}
Here $\Delta_{ijkl}=[\delta_{ik}\delta _{jl}+\delta _{il}\delta _{jk}]$, and
all other cumulants vanish.
In our analysis, the integration over the chaotic
part, given by $h_1(X)$, is defined through a gaussian measure for parametric
random  matrices \cite{caio}
\begin{eqnarray}
P[h_1(X)] &\propto & \exp \left\{ -\frac 12\int
   dXdY\right.  \label{eq:meas} \\
  &&\left.{\rm Tr}\left[ h_1(X)
      {\cal G}^{-1}(X-Y))h_1(Y)\right]\right\}.\nonumber
\end{eqnarray}
The character of the
interaction of the background with the test particle is incorporated into the
correlation function ${\cal G}(X-Y)$, for which we use the form
\cite{brink,bdk1,bdk2}:
\begin{equation}
{\cal G}_{ij}(X)=
\frac{\Gamma ^\downarrow }{2\pi \sqrt{\rho (\varepsilon _i)\rho (\varepsilon
_j)}}
\exp \left [ -\frac{(\varepsilon _i -\varepsilon _j)^2}{2\kappa _0
^2} \right ]G \left (\frac{X}{X_0}\right ).
\end{equation}
This describes a parametric, banded, random matrix where the strength of matrix
elements decreases with increasing level density. Here $G(x)=G(-x)=G^*(x)\le
1$, $G(0)=1$, and the spreading width $\Gamma ^\downarrow$, $\kappa_0$ (linked
with the effective band width $N_0 \approx \kappa _0 \rho(\varepsilon ))$ and
the correlation length $X_0$ are  characteristics of the background.

In order for the measure (\ref{eq:meas}) to be positive definite
$G$ must not decorrelate faster than a gaussian \cite{caio}:
\begin{equation}
G(x) \simeq 1- |x|^\alpha + \cdots,\qquad\qquad \alpha\in (0,2].
\end{equation}
As the position $X$ of
the slow particle changes, the instantaneous energy levels $E_n(X)$
of $[H_b(X)]_{ij}$ change.
Using the above measure, the average fluctuations are
\begin{equation}
\langle\left [ E_n(X) -E_n(Y) \right ] ^2\rangle = D_\alpha |X-Y|^\alpha.
\end{equation}
The energy--spacing fluctuations have a behavior,
which is similar to a L\'evy process
characterized by the diffusion constant $D_\alpha$. The character of these
fluctuations in the eigenvalues $E_n(X)$,
indicated by $\alpha$, will be seen to be related to
L\'evy distributions, which describe the time evolution of the density
matrix for a particle evolving in this chaotic bath.

To develop the dynamical evolution of a free particle evolving in the  presence
of a chaotic background, we take the Hamiltonian of the form:
\begin{equation}
H_{ij}(X,P)=\delta _{ij}\left [ \frac{P^2}{2M}+U(X)\right ] +H_{b,ij}(X).
\end{equation}
The correlated, random--matrix bath can be integrated out in an influence
functional formalism\cite{bdk1}. For our purposes, the $o(\beta)$ action is
sufficient, as well as a weak coupling of the particle to the bath.  
In this case the effective equation for the density matrix of the
test particle has the form:
\begin{eqnarray}
i\hbar \partial _t\rho (X,Y,t) &=&\left\{ \frac{P_X^2}{2M}-\frac{P_Y^2}{2M}
+U(X)-U(Y)\right.  \label{evol} \\
&&-\frac{\beta \Gamma ^{\downarrow }\hbar }{4X_0M}G^{\prime }\left( \frac{%
X-Y }{X_0}\right) (P_X-P_Y)  \nonumber \\
&&\left. +i\Gamma ^{\downarrow }\left[ G\left( \frac{X-Y}{X_0}\right)
-1\right] \right\} \rho (X,Y,t),\nonumber
\end{eqnarray}
where in weak coupling, $G(x)=1-|x|^\alpha$ and
$G'(x)$ above represents $-\alpha{\rm sign}(x) |x|^{\alpha-1}$.

Consider first  a test--particle in the absence of an external field and
interacting with a background with constant average level density ($U(X)=0$
and $\beta =0$). This evolution equation can be solved by passing to the
coordinates $r=(X+X^{\prime })/2$, $s=X-X^{\prime }$. In these variables,
the density matrix has the form:
\begin{eqnarray}
\rho (r,s,t) &=&\int dr^{\prime }\int \frac{dk}{2\pi \hbar }\rho
_0\left ( r^{\prime },s-\frac{kt}{m}\right )
\exp \left[ \frac{ik(r-r^{\prime })}\hbar \right.
\nonumber \\
&&\left. +\frac{\Gamma ^{\downarrow }M}{\hbar k}
\int_{s-kt/M}^sds^{\prime}
\left [ G\left ( \frac{s^{\prime }}X_0 \right ) -1\right ]\right]  \label{rho}
\end{eqnarray}
An initial wavepacket,
$\psi_0(X)=\exp [-X^2/4\sigma^2]/[2\pi \sigma ^2]^{1/4}$,
provides an initial density matrix
$\rho_0(X,X^\prime)=(1/\sqrt{2\pi \sigma ^2})\exp[-(4r^2+s^2)/8\sigma ^2]$.

For the diffusive dynamics of the test particle, we are interested in the
diagonal component of the density matrix $\rho (X,X,t)=\rho (r,s=0,t):$
\begin{eqnarray}
\rho (r,0,t) &=&\int \!\!\int \frac{dr^{\prime }dk}{2\pi \hbar }\rho
_0\left( r^{\prime },-kt/M\right) \exp \left[ \frac{ik(r-r^{\prime })}\hbar
\right.  \nonumber \\
&&\left. -\int_{-kt/M}^0ds^{\prime }\frac{M\Gamma ^{\downarrow }}{k\hbar }
\left| \frac{s^{\prime }}{X_0}\right| ^\alpha \right] \\
&=&\int \frac{dk}{2\pi \hbar }\exp \left[ -k^2\left[ \frac{\sigma ^2}{2\hbar
^2}+\frac{t^2}{8M\sigma ^2}\right] \right.  \nonumber \\
&&\left. -\frac{\Gamma ^{\downarrow }t^{\alpha +1}}{(\alpha +1)\hbar
(MX_0)^\alpha }|k|^\alpha +ik\frac r\hbar \right] .  \label{eq:levy}
\end{eqnarray}
$\rho(X,X,t)$ is nothing more than the spatial probability distribution
$P(X,t)$ for the process. We can now express it in terms of a convolution of
L\'evy distributions:
\begin{equation}\label{eq:px}
\rho (X,X,t)=\int dX^{\prime }{\cal L}_\alpha ^{a(t)}(X^{\prime })
{\cal L}_2^{b(t)}(X-X^{\prime }),
\end{equation}
where
\begin{eqnarray}
 a(t) &=&\frac{\Gamma ^{\downarrow }}{(\alpha +1)\hbar }\left( \frac \hbar {
               MX_0}\right) ^\alpha t^{\alpha +1}, \\
 b(t) &=&\frac{\sigma ^2}2+\frac{\hbar ^2}{8M^2\sigma ^2}t^2.
\end{eqnarray}
As both functions in the integrand of Eq. (\ref{eq:px}) are positive definite,
the spatial  probability $P(X,t)$ is also positive definite. Notice that
the restriction of  $0< \alpha \leq 2$, which came from the short--distance
statistical  correlations in (11) and the requirement of a positive definite
statistical measure, is also the necessary requirement on the
L\'evy distribution
to keep the  resulting time evolution positive definite. Hence the character of
the {\it short--distance} fluctuations is directly responsible for the 
{\it long--time} behavior of the quantum system.

Consider now the short--time and long--time behavior of the dynamics. For
$1<\alpha <2$, in the limit of long times, we expect the $t^{\alpha+1}$ term to
dominate over $t^2$ in (\ref{eq:levy}), so that the density asymptotically
approaches a L\'evy distribution:
\begin{equation}
\rho (X,X,t)\longrightarrow \ a(t)^{-1/\alpha }{\cal L}_\alpha \left(
a(t)^{-1/\alpha }X\right) ,
\end{equation}
while for very short times, the gaussian process is the dominant behavior:
\begin{equation}
\rho (X,X,t)\longrightarrow \ \frac{\sqrt{2}}\sigma \ {\cal L}_2\left( \frac{%
\sqrt{2}}\sigma X\right)
\end{equation}
Specifically, in Eq. (\ref{eq:levy})
the $|k|^\alpha$ term in the exponent dominates
in the long time limit only for momenta $k<k_c$,
where
\begin{equation}
k_c=\left [\frac{a(t)}{b(t)}\right ] ^{\frac{1}{2-\alpha}}
\propto t^{\frac{\alpha -1}{2-\alpha }}.
\end{equation}
For the
special case of $\alpha =2$, the result is gaussian, but the dynamics can be
anomalous, as one can have
turbulent--like diffusion of the type
$\left\langle X^2\right\rangle \sim t^3$\cite{kbd}.  When the level density of
the background is not constant, $\beta>0$ and $\alpha =2$,
one can recover Brownian diffusion\cite{bdk2}. For general $\alpha$ and
$\beta>0$, however, the results are not yet known.
For the range $0<\alpha <1$, the long--time behavior approaches a
gaussian process.
At short--times, the dynamics is influenced by
${\cal L}_\alpha$, and there is a cross--over from short time L\'evy
dynamics to normal gaussian expansion of the wavepacket. In both
cases however, $1<\alpha <2$ and $0<\alpha <1$ respectively, the
second spatial moments are strictly speaking divergent. One should also note
that as neither $a(t)$ nor $b(t)$ are linear in time, even though the
dynamics has the character of a L\'evy process, it is not a L\'evy
stable law.

Efforts to understand unusual stochastic behaviors of dynamical systems has led
to the development of extensions of the Fokker--Planck (FP)
equation\cite{fog,zas,fp}.  These are phenomenological fractional kinetic
equations (restricted to one dimension) in which certain derivatives are
replaced by derivatives of `fractional' order\cite{samko}.
Such approaches have
also found applications in a wide range  of problems from turbulence to
diffusion in porous or viscoelastic media\cite{z1}. We can now explore the type
of stochastic process, which emerges in the  classical limit of our quantum
L\'evy processes, and the connection to multi--dimensional fractional kinetic
theory.

A typical type of phenomenological fractional FP equation has the form
\begin{eqnarray}
\frac{\partial ^\delta P(Q,t)}{\partial t^\beta }&=&\frac{\partial ^\mu }{
\partial (-Q)^\mu }\left( A(Q)P(Q,t)\right) \\
& &+\frac 12\frac{\partial ^{2\nu }}{\partial (-Q)^{2\nu }}\left(
B(Q)P(Q,t)\right). \nonumber
\end{eqnarray}
where $\mu=\nu=1$ in Ref. \cite{fp}, $\mu=\nu$ in Ref. \cite{zas} and
$\delta=\nu=1$ in Ref. \cite{fog}. Here the symbol $\partial^\mu/
\partial x^\mu$ represents the Riemann--Liouville fractional
derivative\cite{samko}, except for Ref. \cite{fog}, where it
represents the Fourier transform of $-k^\mu$.
This equation, while formally constructed, is phenomenological. It is
defined to reproduce anomalous diffusion through scaling formulas such as
$Q^2\sim t^{\gamma}$, where $\gamma$ is a function of $\delta,\mu,\nu$.
A few points should be made here. Generally, the
coefficients $A$ and $B$ are defined  as limits whose existence is postulated
but not known. Further, either the form of the  fractional derivatives is taken
to provide this scaling law, or power law noise is chosen to obtain 
them\cite{fog,bouch}. Such
dynamics  can then be related to L\'evy processes\cite{szk}. Finally, the
extension of these equations to phase space becomes tenuous, since it is not
clear how to include momentum. Not only is it unclear if one should take
fractional derivatives with respect to coordinates, momenta or both, but the
existence of the corresponding coefficients $A$, $B$,... is unknown. Through
our transport equation, we can provide a microscopic interpretation of these
coefficients as well as a systematic manner to construct a fractional kinetic
equation in phase space, whose quantum limit results in L\'evy processes.

To obtain a classical transport
equation, we construct the Wigner transform $f(Q,P,t)$ of the density matrix
$\rho (X,Y,t)$ as
\begin{equation}
f=\int \frac {dR}{2\pi \hbar } \exp\left ( -\frac{iPR}{\hbar}\right )
 \rho \left( Q+\frac R2,Q-
\frac R2,t\right) .
\end{equation}
Applying this to our evolution equation, taking the leading order terms in $
\hbar$, we find
\begin{eqnarray}
&&\frac{\partial f}{\partial t}=\int \frac{dR}{2i\pi \hbar ^2}\exp \left( -
\frac{iPR} \hbar \right) \left\{ -\frac{\hbar ^2}{2M}
\partial _Q\partial _R \right. \\
&& +U\left( Q+\frac R2\right) -U\left( Q-\frac R2\right) -i\Gamma
^{\downarrow }\left| \frac R{X_0}\right| ^\alpha  \nonumber \\
&& \left. +i\gamma \hbar X_0\alpha {\rm sign}(R)
\left| \frac R{X_0}\right|^{\alpha -1}
\partial _R\right\} \rho \left( Q+\frac R2,Q-\frac R2,t\right) .  \nonumber
\end{eqnarray}
This leads naturally to the Reisz fractional integro--differential operator.
This operator, applied to a function $f(P)$, is defined as\cite{samko}
\begin{equation}
(-\Delta_P)^{\frac{\alpha }{2}} f = {\cal F}^{-1} |X|^\alpha {\cal F} f,
\end{equation}
where $\Delta_P$ is the Laplacian (in our case with respect to the momentum
$P$), and ${\cal F}$ represents a Fourier transform.
(This operator is distinct from that proposed in \cite{fog} which did not have
the absolute value, and from \cite{zas}, which uses the Riemann--Liouville
form of this
operator. The Reisz operator is defined as a fractional integral for
Re $\alpha<$0  and as a fractional derivative
for Re $\alpha>$0 through analytic
continuation.) It is convenient to define the operator
$D^\alpha_P=(-i/\hbar)^\alpha (-\Delta_P)^{\alpha/2}$,
since $D^2_P[f]=\partial^2 f/\partial P^2$.
Then the classical limit of our quantum L\'evy process gives rise
to a fractional extension of Kramers equation:
\begin{equation}
\frac{\partial f(Q,P,t)}{\partial t}  + \frac{P}{M}\frac{\partial f(Q,P,t)}{
\partial Q}-\frac{\partial U(Q)}{\partial Q}\frac{\partial f(Q,P,t)}{
\partial P}
\end{equation}
$$
   = \gamma _\alpha \left\{ \overline{D}^{\alpha -1}_P[Pf(Q,P,t)]
    -\frac{2TM}{\alpha\hbar^2}(i\hbar)^\alpha D^\alpha_P[ f(Q,P,t)]\right\},
$$
where $T=1/\beta $ is the temperature and the operator
$\overline{D}^\alpha_P=(-i/\hbar)^\alpha {\cal F}^{-1}
{\rm sign}(X)|X|^\alpha {\cal F}$, with the property
$\overline{D}^1_P[Pf]=\partial (Pf)/\partial P$.
The generalized friction coefficient is given by:
\begin{equation}
\gamma_\alpha=\frac{\beta\Gamma^\downarrow\hbar\alpha}{2MX_0^\alpha}.
\end{equation}
For $\alpha=2$ we recover Kramers equation\cite{van}. What we see is that it is
not the coordinates which acquire the fractional  character, as usually
assumed, but the momenta. Because the coupling to the  background is not
momentum dependent, the correlation function $G(X)$ results only in fractional
derivatives with respect to momenta. This can be traced back to the nature of
the chaotic correlations in Eq. (\ref{evol}). Further, these  processes,
related to  L\'evy processes, do not require the introduction of  fractional
time derivatives.  We note here that our transport theory has a consistent
classical limit for  all of these transport coefficients only when they remain
finite as $\hbar  \rightarrow 0$. This requires in turn that the parameters of
our quantum theory  cannot remain constant as $\hbar \rightarrow 0$, if we are to
recover a well defined classical transport. Finally, we observe that this
approach provides finite coefficients $D_{QQ}$, $D_{PP}$, $D_{QP}$ and so forth
(e.g. A, B,...)  for a  fractional kinetic equation in phase space.

We have shown that the quantum evolution of a wavepacket in a chaotic
environment can lead to reduced density matrices which behave as
L\'evy processes. The short distance energy
fluctuations of the background,  which are characterized by a parameter
$\alpha\in(0,2]$, are found to be precisely related to the quantum time
evolution with a L\'evy process of the same character $\alpha$.  For
$\alpha=2$ one has gaussian processes, which can display normal to
turbulent--like diffusion or Brownian diffusion ($\beta>0$), while for
$\alpha=1$ one has the dynamics of the Dyson process. The general quantum
evolution of a wavepacket displays a cross--over between gaussian and L\'evy
dynamics. In passing to the classical limit of this behavior, we find that the
dynamical evolution  results in a fractional kinetic equation, which is a
generalization of Kramers equation. For $\alpha=2$ Kramers theory is recovered.
This approach provides a means to develop fractional kinetic theory in more
than one dimension, since the expansion coefficients are determined from the
microscopic theory. It also provides the possibility to explore the connections
between quantum and classical transport in chaotic systems, as well as the
links between chaos, quantum statistical  fluctuations, L\'evy processes and
classical fractional dynamics.

\end{document}